\documentstyle{article}

\newlength{\defaultparindent}
\newenvironment{Default Paragraph Font}{}{}
\input{tcilatex}

\begin{document}

\begin{center}
Model of Expansive Nondecelerative Universe and Unified Approach \

to Fundamental Physical Interactions\bigskip

Miroslav Sukenik, Jozef Sima

Slovak Technical University, Radlinskeho 9, 812 37 Bratislava, Slovakia

e-mail: sima@chelin.chtf.stuba.sk\medskip

and Julius Vanko

Comenius University, Dep.Nucl.Physics, Mlynska dolina F1, 842 48
Bratislava,
Slovakia

e-mail: vanko@fmph.uniba.sk\bigskip
\end{center}

Abstract. The contribution provides the background of the model of
Expansive
Nondecelerative Universe, rationalizes the introduction of Vaidya
metrics
allowing thus to localize and quantify gravitational energy. A unifying
explanation of the fundamental physical interactions is accompanied by
demonstration of some consequences and predictions relating to
cosmological
problems.\bigskip

I. Introduction\medskip

Starting from the beginning of 80's, an inflation model of the universe
acquired a dominant position in cosmology. Of the main designers of the
model, A. Guth and A. Linde should be mentioned. The model has been able
to
eliminate certain cosmological problems such as a problem of the
universe
flatness and horison. At the same time it has open, however, several new

questions. There are still significant differences in the calculated
values
of universe age depending on cosmological theories applied. Inflation
model
has not been able to precise some important parameters of the universe
such
as Hubble's constant or deceleration parameter. It has brought no
important
knowledge on understanding the gravitation and its relation to the other

physical interactions. Moreover, in accordance with some analyses [1],
the
initial nonhomogenities should not be eliminated but they are rather
enhanced within the inflation period.

The mentioned open questions have been a challenge for developing other
models of the Universe, one of them being our Expansive Nondecelerative
Universe (ENU) model [2]. Here the background of the ENU model is given
and
its advantages in solving some problems are documented.

It should be pointed out that both the inflation and ENU models exhibit
some
common features, such as an increase in the universe mass. They differ
in a
way of the increase. In the inflation model the mass increase is a
consequence of its emerging beyond the causal horison due to a
deceleration
of the Universe expansion caused by gravitational forces. In the ENU
model
the matter is created simultaneously with the gravitational energy that
is
negative. The total mass-energy is thus constant and equal to zero which
is
in accordance with the laws of conservation. Such an universe can
permanently expand with the velocity of light. One of the key
differences of
the inflation and ENU models concerns the red shift that is constant in
ENU
but decreasing in the inflation model.\bigskip

II. Theoretical background\medskip

Applying Robertson-Walker metrics to Einstein's field equations [3],
Friedmann's [4] equations of the universe dynamics are obtained\bigskip

$\left( \frac{da}{dt}\right) ^{2}=\frac{8}{3}\pi .G.\rho
.a^{2}-k.c^{2}+%
\frac{1}{3}\Lambda .a^{2}.c^{2}$ \hfill (1)\bigskip

$2a.\left( \frac{d^{2}a}{dt^{2}}\right) +\left( \frac{da}{dt}\right)
^{2}=-%
\frac{8\pi .G.p.a^{2}}{c^{2}}-k.c^{2}+\Lambda .a^{2}.c^{2}$ \hfill
(2)\bigskip

where $a$ is the gauge factor, $\rho $ is the mean density of the
universe,

$k$ is the curvature index, $\Lambda $ is the cosmological member, and
$p$
is the pressure. These corner-stones of ENU are defined as
follows\bigskip

$a=c.t_{c}$ \hfill (3)\bigskip

where $t_{c}$ is the cosmological time,\bigskip

$k=0$ \hfill (4)\bigskip

$\Lambda =0$ \hfill (5)\bigskip

Introducing equations (3), (4) and (5) into (1) and (2) has led [4]
to\bigskip

$c^{2}=\frac{8\pi .G.\rho .a^{2}}{3}=-\frac{8\pi .G.p.a^{2}}{c^{2}}$
\hfill
(6)\bigskip

and at the same time\bigskip

$\epsilon =\rho .c^{2}$ \hfill (7)\bigskip

Relations (6) and (7) lead directly to the state equation\bigskip

$p=-\frac{\epsilon }{3}$ \hfill (8)\bigskip

The negative sign of the right side documents the negative value of the
gravitational energy. Equation (6) can be rewritten as follows\bigskip

$\rho =\frac{3c^{2}}{8\pi .G.a^{2}}$ \hfill (9)\bigskip

For a flat universe it holds\bigskip

$\rho =\frac{3m_{U}}{4\pi .a^{3}}$ \hfill (10)\bigskip

where $m_{U}$ is the mass of the universe matter. A comparison (9) and
(10)
gives\bigskip

$a=c.t_{c}=\frac{2G.m_{U}}{c^{2}}$ \hfill (11)\bigskip

that manifests the matter creation in time.

Within the first approximation, the density of the gravitational field
is
described by Tolman's relation (12)\bigskip

$\epsilon _{g}=-\frac{R.c^{4}}{8\pi .G}$ \hfill (12)\bigskip

in which $R$ denotes the scalar curvature. Contrary to a more frequently

used Schwarzschild metrics (in which $\epsilon _{g}=0$ outside a body
and $%
R=0$ which prevents from localization of the gravitational energy), in
Vaidya metrics [5] $R\neq 0$ and $\epsilon _{g}$ may thus be quantified
and
localized also outside a body. In ENU it holds also\bigskip

$\frac{dm}{dt}=\frac{m}{t_{c}}$ \hfill (13)\bigskip

Using Vaidya metrics and (13), an interrelationship (14) of scalar
curvature
and gravitational diameter $r_{g(m)}$ of a body with the mass\bigskip

$m$ emerges [6]\bigskip

$R=\frac{6G}{r^{2}.c^{3}}\left( \frac{dm}{dt}\right) =\frac{6G.m}{%
t_{c}.r^{2}.c^{3}}=\frac{3r_{g(m)}}{a.r^{2}}$ \hfill (14)\bigskip

Introducing (14) into (12) the density of energy $\epsilon _{g}$ induced
by
such a body at a distance $r$ can be expressed as:\bigskip

$\epsilon _{g}=-\frac{3m.c^{2}}{4\pi .a.r^{2}}$ \hfill (15)\bigskip

Equation (15) can be rewritten as\bigskip

$\epsilon _{g}=\frac{3E_{g}}{4\pi .\lambda ^{3}}$ \hfill (16)\bigskip

where $E_{g} $  is the energy of a quantum of gravitational field and

$\lambda $ its Compton's wavelength\bigskip

$\lambda =\frac{\hbar .c}{E_{g}}$ \hfill (17)\bigskip

Substituting $\lambda $ in (17) for (16) and comparing the result with
(15),
the expression for an energy quantum is obtained:\bigskip

$\left| {}\right. E_{g}\left. {}\right| =\left( \frac{m.\hbar
^{3}.c^{5}}{%
a.r^{2}}\right) ^{1/4}$ \hfill (18)\bigskip

where $E_{g} $  is the quantum of gravitational energy created by a body

with the mass $m$  at a distance $r$ . Relation (18) is in conformity
with
the limiting values: the maximum energy is represented by the Planck's
energy, the minimum energy equals the energy of a photon with the
wavelength
identical to the universe dimension ( $a=\lambda $ ) .

Gravitational output, defined as an amount of the gravitational energy
emitted by a body with the mass $m$ per unit time unit can be derived
from
(15) as\bigskip

$P_{g}=\;\frac{d}{dt}\int \epsilon _{g}.dV\;\cong
\;-\frac{m.c^{3}}{a}\;=-%
\frac{m.c^{2}}{t_{c}}$ \hfill (19)\bigskip

Gravitational force is a far-reaching force with ostensibly unlimited
range.
Due to the existence of hierarchic rotational gravitational systems, the

range is, however, actually finite. This is a reason for introducing so
called ``effective gravitational range`` $r_{ef(g)}$ , i.e. the distance
at
which the density of gravitational field of a given body is equal to
critical density [7]. It follows from (9) and (15) that\bigskip

$\frac{3c^{4}}{8\pi .G.a^{2}}=\frac{3m.c^{2}}{4\pi .a.r^{2}}$ \hfill
(20)\bigskip

and, in turn\bigskip

$r=r_{ef(g)}=\left( r_{g(m)}.a\right) ^{1/2}$ \hfill (21)\bigskip

where $r_{ef(g)}$ is the effective gravitational range of a body with
the
gravitational radius $r_{g(m)}$ . The present value of the gauge factor
is\bigskip

$a\cong 1.3x10^{26}m$ \hfill (22)\bigskip

Based on the above rationalization, it is possible to determine the
lightest
particle able to exerts gravitational influence on its surroundings. The

partice has the mass $m_{x}$ and its gravitational range is identical to
its
Compton's wavelength. Stemming from the following relation\bigskip

$\left( \frac{2G.m_{x}.a}{c^{2}}\right) ^{1/2}=\frac{\hbar }{m_{x}.c}$
\hfill (23)\bigskip

the mass of the particle is\bigskip

$m_{x}\cong \left( \frac{\hbar ^{2}}{2G.a}\right) ^{1/3}\cong
10^{-28}kg$
\hfill (24)\bigskip

and its Compton's wavelength\bigskip

$\lambda _{x}=\frac{\hbar }{m_{x}.c}\cong 10^{-15}m$ \hfill (25)\bigskip

In our paper [7] we discussed the advantages and consequences of
introduction of $m_{x} $  into the dimensionless gravitational constant

$\alpha _{g}.$ Relation (21) allows to determine an amount of dark
matter if
dimensions of the radiation emitting matter of a corresponding
hierarchic
rotational gravitational system (galaxies, clusters and superclusters)
are
known [8].\bigskip

III. Unification of gravitational and electromagnetic
interactions\medskip

Field equations of the geometrized electrodynamics can be written as
follows
[9]\bigskip

$R_{ik}-\frac{g_{ik}.R}{2}=\frac{8\pi .e}{m.c^{4}}T_{ik}$ \hfill
(26)\bigskip

where $T_{ik}$ is the momentum-energy tensor expressed by means the
density
of electromagnetic energy $\rho _{e}.c^{2\bigskip }$

$T_{ik}=\rho _{e}.c^{2}.u_{i}.u_{k}$ \hfill (27)\bigskip

As a metrics of the field of a charged particle, Rieman metrics can be
used
in the form [9]\bigskip

$ds^{2}=\left( 1-\frac{r_{e}}{r}\right) c^{2}.dt^{2}-\left(
1-\frac{r_{e}}{r}%
\right) ^{-1}dr^{2}-r^{2}\left( d\theta ^{2}+\sin ^{2}\theta .d\varphi
^{2}\right) $ \hfill (28)\bigskip

where $r_{e}$ is the electromagnetic radius\bigskip

$r_{e}=\frac{2e^{2}}{m.c^{2}}$ \hfill (29)\bigskip

Monopole radiation of a charged body can be described by Vaidya metrics
[9]
as\bigskip

$ds^{2}=\left( 1-\frac{2e}{m.r.c^{2}}.\frac{de}{dt}\right)
c^{2}.dt^{2}-\left( 1+\frac{2e}{m.r.c^{2}}.\frac{de}{dt}\right) \left(
d^{2}x+d^{2}y+d^{2}z\right) $ \hfill (30)\bigskip

Density of electromagnetic field energy $\epsilon _{e}$ obtained from
(26)
can be written as\bigskip

$\epsilon _{e}=-\frac{m.c^{4}.R}{8\pi .e}$ \hfill (31)\bigskip

where $R$ is the scalar curvature that, using (30), adopted the
form\bigskip

$R=\frac{6}{r^{2}.c^{3}}.\frac{de}{dt}=\frac{6e}{a.r^{2}.c^{2}}$ \hfill
(32)\bigskip

Introducing (32) into (31), relation (33) identical to (15) is obtained
and
this identity can be taken as an evidence of the unity of gravitational
and
electromagnetic interactions.\bigskip

$\epsilon _{e}=-\frac{3m.c^{2}}{4\pi .a.r^{2}}$ \hfill (33)\bigskip

In the initial period of the universe expansion $(t\approx 10^{-44}
s;T\approx 10^{32} K)$  all the fundamental interactions were unified.
Two
factors are worth keeping in mind when evaluating unification of the
interactions:

1) energy - mass at the time of unification/separation,

2) range typical for a given kind of interaction.

The value of energy-mass is identical for unification of all physical
interactions. It holds ( $m_{Pc}$ is the Planck's mass)\bigskip

$m_{g}=m_{Pc}=\left( \frac{\hbar .c}{G}\right) ^{1/2}\cong 10^{19}GeV$
\hfill (34)\bigskip

Since both gravitational and electromagnetic interactions are of
far-reaching nature, a question of dimension limit does not play any
important role in their unification.\bigskip

IV. Unification of gravitational and strong interactions\medskip

Binding energy $E_{s}$ of two quarks increases with the square of their
distance. This phenomenon is expressed in [9 - 11] as follows\bigskip

$E_{s}=\frac{\hbar .c}{a}.\frac{r^{2}}{l_{Pc}^{2}}$ \hfill (35)\bigskip

where $l_{Pc}$ is the Planck's distance\bigskip

$l_{Pc}=\left( \frac{G.\hbar }{c^{3}}\right) ^{1/2}\cong 10^{-35}m$
\hfill
(36)\bigskip

There are authors proposing a linear dependence of $E_{s}$ on $r$ .
However,
relation (35) is in good accordance with asymptotic freedom for limiting

cases. When\bigskip

$r=l_{Pc}$ \hfill (37)\bigskip

the energy reaches the minimum possible value\bigskip

$E_{s}=\frac{\hbar .c}{a}$ \hfill (38)\bigskip

i. e. the energy of a photon with the maximum wavelength $a.$ On the
other
hand, if\bigskip

$r=a$ \hfill (39)\bigskip

then\bigskip

$E_{s}=m_{U}.c^{2}$ \hfill (40)\bigskip

that represents the maximum energy. At a typical distance\bigskip

$r\cong 10^{-15}m$ \hfill (41)\bigskip

the energy value is\bigskip

$E_{s}\cong 10^{-11}J$ \hfill (42)\bigskip

This value is very close to the kinetic energy of $\pi ^{+} $  mesons
(200
MeV) at their scattering on protons followed by the formation of
resonance.

Unification of gravitational and strong interactions is conditioned by
the
following equality\bigskip

$\frac{G.m_{g}^{2}}{r}=\frac{\hbar .c}{a}.\frac{r^{2}}{l_{Pc}^{2}}$
\hfill
(43)\bigskip

It stems from (34) and (43) that\bigskip

$r=\lambda _{x}=\frac{\hbar }{m_{x}.c}\cong 10^{-15}m$ \hfill
(44)\bigskip

The value of $m_{x} $  is given in (24). Calculated limiting distance
(44)
actually correspond to a real range of the nuclear forces that, in turn,
can
be understood as an evidence of unification of the gravitational and
strong
interactions.

Based on (24), (35) and (44), a time dependent decrease in the energy of

strong interactions can be obtained as\bigskip

$E_{s}\approx a^{-1/3}\approx t_{c}^{-1/3}$ \hfill (45)\bigskip

At present, the typical value of this energy is about 100 MeV. The
magnitude
of binding energy of an electron in atom is 1 eV. Based on (45), the
values
of nuclear strong interaction and electron binding energies should
approach
in time\bigskip

$t\cong 10^{34}$ years\hfill (46)\bigskip

This result of ENU is in good agreement with the time of baryonic matter

disintegration predicted by the GUT theory.\bigskip

V. Unification of gravitational and weak interactions\medskip

The cross section $\sigma $ of weak interaction can be expressed as [12,

13]\bigskip

$\sigma \cong \frac{g_{F}^{2}.E_{w}^{2}}{(\hbar .c)^{4}}$ \hfill
(47)\bigskip

where $g_{F}$ is the Fermi's constant, $E_{w}$ is the energy of weak
interactions that, based on (47), can be formulated by relation\bigskip

$E_{w}\cong \frac{r.\hbar ^{2}.c^{2}}{g_{F}}$ \hfill (48)\bigskip

where $r$ represents the effective range of weak interactions related in
a
limiting case to the mass $m_{ZW}$ of vector bosons Z and W\bigskip

$r=\frac{\hbar }{m_{ZW}.c}$ \hfill (49)\bigskip

The maximum energy of weak interaction obtained from (48) and (49) is
then
given as\bigskip

$E_{w,\max }\cong m_{ZW}.c^{2}$ \hfill (50)\bigskip

Equations (48), (49) and (50) lead to an expression for the mass of the
bosons Z and W\bigskip

$m_{ZW}^{2}\cong \frac{\hbar ^{3}}{g_{F}.c}\cong \left( 100GeV\right)
^{2}$
\hfill (51)\bigskip

giving the value that is in good agreement with the known actual value.

When unifying gravitational and weak interactions, it must hold\bigskip

$\frac{G.m_{g}^{2}}{r}=\frac{\hbar ^{2}.c^{2}.r}{g_{F}}$ \hfill
(52)\bigskip

Using (34) and (52) the value of limiting range of weak interactions is
obtained\bigskip

$r=\lambda _{ZW}=\frac{\hbar }{m_{ZW}.c}\cong 10^{-18}m$ \hfill
(53)\bigskip

that proves the unity of gravitational and weak interactions.\bigskip

VI. Formation of solar antineutrinos from the viewpoint of ENU\medskip

Stemming from one of the most significant achievements of the ENU model
-
localization of gravitational energy - and based on the knowledge on
weak
interactions some consequences related to the Sun occurring processes
can be
drawn. It should be pointed out that the essence of the following
discussion
is to be taken as a scientific hypothesis rather than the definite
conclusions.

The gravitational energy output of a body with the mass $m$ is
established
by relation (19). This energy is negative. Due to the conservation laws
validity, a corresponding amount of energy-matter must be created. In
the
case of the Sun it represents\bigskip

$P=\frac{m_{S}.c^{2}}{t_{c}}$ \hfill (54)\bigskip

where $m_{S} $  is the mass of the Sun.

Due to the chemical composition of the Sun it appears justifiable to
assume
that a significant part of the matter created are represented by
neutrons.
Owing to the weak interactions, neutrons undergo a decay to protons,
electrons and antineutrinos:\bigskip

$n\rightarrow p^{+}+e^{-}+\nu $ \hfill (55)\bigskip

Based on the fact that the rest mass of the neutron exceeds the proton
mass
about on the mass of two electrons and the mass of antineutrinos (as
well as
neutrinos) can be omitted, analysing relation (55) in a more detail, the
law
of energy conservation leads to a conclusion that each process (55) is
accompanied by an energy release of about 0.5 MeV. Taking the whole Sun
into
account, the energy output totalled to\bigskip

$P=\frac{m_{S}}{t_{c}.m_{n}}.m_{e}.c^{2}\cong 2x10^{26}W$ \hfill
(56)\bigskip

where $m_{n}$ and $m_{e}$ are the mass of neutron and electron,
respectively. The result in (56) is in excellent agreement with the
assessed
actual radiation energy output that is\bigskip

$P\cong 4x10^{26}W$ \hfill (57)\bigskip

A number of antineutrinos being formed per 1 second via the neutron
decay
(55) is given as\bigskip

$N_{\nu }=\frac{m_{S}}{t_{c}.m_{n}}\cong 10^{39}s^{-1}$ \hfill
(58)\bigskip

It is worth emphasizing that the neutrinos/antineutrinos formation can
occur
both inside the Sun and in its environment (solar corona).

The present hypothesis casts in no case any doubt on the existence of
thermonuclear processes inside the Sun. At the given conditions they
must
proceeds. It is, however, possible that simultaneously also the creation

occurrs and its effect may be related to the known deficit of solar
neutrinos. In addition, the question concerning the solar corona
temperature
can be answered from the presented viewpoint.\bigskip

VII. ENU model and wave function of the universe\medskip

Based on (18) when\bigskip

$m=m_{U}$ \hfill (59)\bigskip

$r=a$ \hfill (60)\bigskip

the energy of gravitational quanta of the universe is obtained
[12]\bigskip

$\left| E_{g}\right| =\left( \frac{\hbar ^{3}.c^{7}}{G.a^{2}}\right)
^{1/4}$
\hfill (61)\bigskip

Expressing (61) by wave function\bigskip

$E_{g}.\Psi _{g}=i.\hbar .\frac{d\Psi _{g}}{dt}$ \hfill (62)\bigskip

where\bigskip

$\Psi _{g}=e^{-i.\omega .t}$ \hfill (63)\bigskip

and at the same time\bigskip

$\omega =\frac{c}{(l_{Pc}.a)^{1/2}}=(t_{Pc}.t_{c})^{-1/2}$ \hfill
(64)\bigskip

where $t_{Pc}$ is the Planck's time. It follows from (63) and (64)
that\bigskip

$\Psi _{g}=e^{-i(t_{Pc}.t_{c})^{-1/2}.t}$ \hfill (65)\bigskip

i.e. the wavefunction is dependent on cosmological time. Further, the
expansion of the Universe is associated with a decrease in the frequency
of
gravitational waves. The rate of the expansion is in ENU constant, wave
function $\Psi _{g} $  will vary due to the uncertainty principle. Such
fluctuations must cause fluctuations of relict radiation. The mentioned
nonhomogenities will increase in time and have given rise to the present

galaxies, clusters and superclusters.

The experimental value of the fluctuation based anisotropy $A$
is\bigskip

$A=\frac{\Delta T}{T}\cong 10^{-5}$ \hfill (66)\bigskip

Given the interrelationship between temperature and energy, it must
hold\bigskip

$\Delta E\cong 10^{-5}E$ \hfill (67)\bigskip

and if uncertainty of time is identified with cosmological time\bigskip

$\Delta t=t_{c}$ \hfill (68)\bigskip

then the Heisenberg's equation of uncertainty can be, based on (66),
(67)
and (68), written as\bigskip

$10^{-5}E.t_{c}\cong \hbar $ \hfill (69)\bigskip

and relations (62), (65) and (69) lead to the time when fluctuations or
anisotropy appeared and developed\bigskip

$t_{c(A)}\cong 10^{10}t_{Pc}\cong 10^{-33}s$ \hfill (70)\bigskip

Since in the primordial period of the universe\bigskip

$T\approx a^{-1/2}\approx t_{c}^{-1/2}$ \hfill (71)\bigskip

in the time about $10^{-33}$ s, temperature of the universe based on ENU

was\bigskip

$T\cong 10^{26}K$ \hfill (72)\bigskip

which is in excellent agreement with the value calculated in [14].

If all the above rationalizations stemming from ENU model are correct,
then
the equivalence (73) must hold.\bigskip

$\left| \Sigma E_{g}\right| =m_{U}.c^{2}$ \hfill (73)\bigskip

Further, the validity of (73) will be checked. Hawking found [15] a
close
relationship between gravitation and thermodynamics. Using
Stefan-Boltzmann
law the following relation can be written\bigskip

$\int \frac{4\sigma .T^{4}}{c}dV\cong M_{U}.c^{2}$ \hfill (74)\bigskip

Substitutions\bigskip

$T\cong \frac{\left| E_{g}\right| }{k}$ \hfill (75)\bigskip

$k^{4}\cong \sigma .\hbar ^{3}.c^{2}$ \hfill (76)\bigskip

and the subsequent integration over whole volume of the universe (that
is
confined by a causal horizon) taken (11) into account leads directly to
(73).

Similarly the validity of relation\bigskip

$\left| \epsilon _{g}\right| =\epsilon _{cr}$ \hfill (77)\bigskip

where $\epsilon _{cr}$ is the critical density of the universe matter
can be
proved.\bigskip

VIII. Time of the anisotropy creation\medskip

Probably it is not a pure coincidence that the gravitational influence
of
vector bosons X and Y started just at at the time of anisotropy
creation\medskip

$t_{c(A)}\cong 10^{-33}s$ \hfill (78)\bigskip

Because of the mass of hypothetical particles mentioned in (24) is just
of
the order $10^{15}-10^{16}$ GeV it may be concluded that their
appearance
and, in turn, their gravitational impact are the very reasons having
caused
both the value of the anisotropy and the time of its creation. If the
introduction of anisotropy $A$ (66) is followed by a subsequent
modification
of (70) into the form\bigskip

$a_{(A)}\cong \frac{l_{Pc}}{A^{2}}$ \hfill (79)\bigskip

where $a_{(A)}$ is the gauge factor at a cosmological time $t_{c(A)}$
from
(78), then (24) is transformed to\bigskip

$a_{(A)}\cong \frac{\hbar ^{2}}{G.m_{XY}^{3}}$ \hfill (80)\bigskip

where $m_{XY}$ is the mass of vector bosons X and Y expressed from (79)
and
(80) as\bigskip

$m_{XY}\cong (A)^{2/3}.m_{Pc}$ \hfill (81)\bigskip

Thus a degree of accuracy in an anisotropy $A$ measurement will help to
determine the mass of the vector bosons X and Y.\bigskip

IX. Conclusions\medskip

It is becoming still more and more obvious that solution of several
problems
of cosmology and astrophysics depends of the ability of researchers to
localize gravitational energy. In this area three main streams of
opinions
can be identified: 1) gravitational field energy is localizable but a
corresponding ``magic`` formula for its density is to be found; 2)
gravitational energy is nonlocalizable in principle; 3) gravitational
energy
does not exist at all since the gravitational field is a pure geometric
phenomenon. It seems that solution of this enigma lies in metrics
applied.
It has stemmed from the starting points of the ENU model that a metrics
involving changes in matter due to its permanent creation must be used.
Vaidya metrics [5] has met the requirements of the ENU model and its
utilization allowed to offer answers to many questions, explain some
known
facts in independent ways, correct some opinions or demostrate their
limitations and predict deep mutual interrelationships of natural
phenomena.
In addition to theoretical approach present in this contribution, we
proposed a procedure for the experimental determination of gravitational

field energy [16]. Based on the up-to-date knowledge we are convinced
that
the time is ripe to accept the idea of localizability of gravitational
field
energy, develop adequate mathematic tools and take advantage this
opportunity to look at old problems from new viewpoints.\bigskip

References\medskip

1..R. Penrose, The large, the small and the human mind, Cambridge
University
Press, Cambridge, 1997, p. 44

2..V. Skalsky, M. Sukenik, Astrophys. Space Sci., 178 (1991) 169

3..A. Einstein, Sitzb. Preuss. Akad. Wiss., 45 (1915) 844

4..A.A. Friedmann, Z. Phys., 10 (1922) 377; 21 (1924) 326

5..P.C. Vaidya, Proc. Indian Acad. Sci., A33 (1951) 264

6..J. Sima, M. Sukenik, General Relativity and Quantum Cosmology,
Preprint:
gr-qc/9903090

7..J. Sima, M. Sukenik, M. Sukenikova, General Relativity and Quantum
Cosmology, Preprint: gr-qc/9910094

8..V. Skalsky, M. Sukenik, General Relativity and Quantum Cosmology,
Preprint: 9603009

9..G.I. Shipov, The theory of physical vacuum, NT Center, Moscow, 1993,
p.
10, 105, 113, 166 and 183 (in Russian)

10..V. Skalsky, M. Sukenik, Astrophys. Space Sci. 236 (1996) 295

11..M. Sukenik, J. Sima, General Relativity and Quantum Cosmology,
Preprint:
gr-qc/9911067

12..I.L. Rosenthal, Adv. Math. Phys. Astr., 31 (1986) 241 (in Czech)

13..M. Sukenik, J. Sima, J. Vanko, General Relativity and Quantum
Cosmology,
Preprint: gr-qc/0001059

14..S.W. Hawking, R. Penrose, The nature of space and time, Princeton
University Press, Princeton, 1996

15..S.W. Hawking, Commun. Math. Phys., 43 (1975) 199

16..M. Sukenik, J. Sima, J. Vanko, General Relativity and Quantum
Cosmology,
Preprint: gr-qc/0006068

\end{document}